# Combining Structured Corporate Data and Document Content to Improve Expertise Finding

Alistair McLean, Anne-Marie Vercoustre, MingFang Wu
CSIRO – ICT Centre
Clayton, Victoria, Australia

Ph. (+61) 03 9545 8461, email: {firstname}.{lastname}@csiro.au

## ABSTRACT

In this paper, we present an algorithm for automatically building expertise evidence for finding experts within an organization by combining structured corporate information with different content. We also describe our test data collection and our evaluation method.

Evaluation of the algorithm shows that using organizational structure leads to a significant improvement in the precision of finding an expert. Furthermore we evaluate the impact of using different data sources on the quality of the results and conclude that Expert Finding is not a "one engine fits all" solution. It requires an analysis of the information space into which a solution will be placed and the appropriate selection and weighting scheme of the data sources.

## Categories and Subject Descriptors

H3.3 Information Search and Retrieval

## 1. INTRODUCTION

The main source of competitive advantage of an organization is its knowledge assets and its learning capacity [1]. Traditional document centric knowledge management approaches have mostly focused on capturing relevant corporate documents and making them available through search or more proactive delivery mechanisms. These approaches have failed in making the knowledge in the corporate memory operationally available (e.g. making the information relevant to the task at hand and converting it into effective actions).

We argue that finding knowledge in an organization is not just about finding documents, but often finding the right person. Indeed Stobie [18] recommends that organizations embarking on any knowledge management program should start by focusing on what counts "to make better use of the expertise in an organization by putting those who know in touch with those who want to know".

Looking at the problem from a different angle, [17] report that expert finding in organizations is done with one or both of the following goals in mind: (1) to find someone as a source of information and (2) to find someone who can perform a given organizational or social function. They relate these goals to "information need" and "expertise need".

Both Snowden [16] and The McKinsey quarterly article [9] also note that knowledge is contextual and thus systems need to have at their disposal many different facets of expertise information in order to be able to provide the most suitable answer to a query --- this also implies that knowing something about the query is important too.

In the corporate world, needs for finding experts include:

- For workers to identify people to ask for expert opinions and perhaps collaborate
- To be able to get a good picture of capabilities and areas of expertise
- To be able to identify appropriate people to build teams
- Develop a sense of community by finding people with similar expertise, or by accessing expertise information about a newly met person.

As noted above, there are many different aspects and needs for expertise finding. We are developing PeopleFinder, a tool for finding people in an enterprise based on their expertise. Unlike some commercially available tools our approach does not rely on a profile database but on a combination of the documents that people create and publish during their actual work and organizational knowledge encoded as an organization's structure. This paper examines whether combining structured corporate information with unstructured content leads to an improvement in the precision of finding an expert within an organization. In order to do this we have developed a novel expertise finding algorithm and a test data collection that allows us to determine that this algorithm improves performance.

## 2. RELATED WORK

There are essentially three approaches to finding experts: 1) a database approach that stores a set of skills, often rated, for each person, and 2) evidence-based approaches that compute a profile for a person according to "electronic evidence" and 3) a referral, or social network approach that links people through a network.

The first approach is often extended by some form of taxonomy mechanism that allows for a fixed vocabulary and greater usability when searching and browsing. A taxonomy mechanism is not fixed to the skills database approach however, as document classification can be used to attach taxonomic categories to people given their evidence-based profile.

This approach has most often been used in organizations where peoples' skills are assessed and recorded in a skills database (eg Microsoft's SPUD [7] or SAGE's PeopleFinder at http://sage.fiu.edu). However, in large organizations, particularly those spread over multiple offices, undergoing organizational change or experiencing high staff turnover, it can be difficult to keep track of employee expertise. Furthermore a fixed set of terms are used to describe skills which leads to terminological mismatches.

Other work such as Peer Helper and more recently I-Help [8][13] use a knowledge profile of a helper which is organized to particular tasks a user may be engaged in. The profile indicates, among other things, ability, willingness and availability for that person to help with a task (or step in a task). This information is gathered using an initial questionnaire, learner models and tracking of actual help given for a task against the learner model and feedback given from the user requesting the help. Other systems such as expert marketplace (www.experts-exchange.com) rely purely on a scoring mechanism whereby the person asking for help scores the person who answers his question. Neither of these approaches is particularly suitable to an organization where minimal effort is required to keep the assessment of knowledge up to date and where we are trying to capture the expertise of people who might not actively take part in voluntary "market places".

[3] describes and analyses two expertise finding systems developed by NASA. The first, SAGE, uses various educational databases that include fields of expertise, names, funding agencies and universities. It uses full text indexing as well as a thesaurus of concepts. Search is based on a recognized valid indicator of expertise based on the funded research grants received. The Expert Locator at the Kennedy Space Center, locate experts based on published documents by extracting names (using a name finding algorithm) and skill descriptions. Experts are ranked according to the number of relevant documents that contain their name (not according to the relevance of the documents) and a taxonomy is used to expend queries. Other work, e.g. [5], take the notion of taxonomies further and develop an entire skills ontology.

The second approach often uses the content of electronic forms of communication (reports, publications, emails, etc) that people write as evidence of their expertise. This can be complemented with the more formal structured data found in corporate databases such as human resource databases, project reporting databases [14]. This approach is closest to ours, but we also differentiate between the type of the data and take advantage of information encoded in an organization's structure.

Finally, there are "interest-based" mechanisms that examine what people are looking at as well as what they are writing [2] [11]. A further mechanism [4] extends this to look at the navigational paths to web documents taken by experts.

The ExpertFinder system of MITRE [12] identifies experts from heterogeneous documents such as those published by employees themselves (eg. resume, homepage, technical papers and presentation) and project descriptions, newsletters and announcements. These documents are then indexed and searched like other search engines. The ExpertFinder system considers someone an expert in a particular topic if they are linked to a wide range of documents and/or a large number of documents about that topic. This system was evaluated on five specialty areas and achieved 41% precision on average at top five retrieved experts.

[19] focuses on finding experts to help in Java programming. Java programs are used as evidence and the domain is modeled using java class hierarchies and relations. User models are built from the classes they have used in their programs and the frequency of use. The system can recommend experts to solve a problem (defined by free text query or a selection of classes). The recommended expert will have expertise that best matches the query and is closest (above) the user's own expertise. Personal profiles can been displayed and updated manually.

In [20] the user's expertise on a given subject is related to the importance of the related pages he has accessed. The importance of the page is a function of the number of links to this page and the number of (important) users who have accessed this page. They propose an iterative algorithm to measure the importance of pages and users. Taking into account documents that the expert has accessed raises the issue of privacy. In contrast, our algorithm ranks experts according to important pages or documents that he/she has created, or that are linked to those important pages. An important page in our system is a page that likely to contain information related to the expert's skills, such as his home page, or the descriptions of projects he has been involved in.

The third approach creates the network through analysis of electronic evidence such as author co-occurrence in papers or analysis of communications traffic.

In [10] the authors postulate that the best way of finding an expert is through what is called referral chaining whereby a seeker finds the needed expert through referral by colleagues. The system uses co-occurrence of names in documents as evidence of relationships. See also [15] where an analysis of email logs calculates distances between any two people.

## 3. PEOPLE-FINDER ARCHITECTURE

CSIRO's original prototype, P@NOPTIC Expert [6] is a web based system which automatically identifies experts in an area, based on the documents already published on an organization's intranet. Like a standard web search engine the system takes a subject query and returns a list of experts. The limitations of this initial prototype included, in some instances, low quality results due to poor quality documents being used as expertise evidence. In the work presented in this paper we refine the computation of a person's expertise by (a) accepting more documents and documents that may not contain their name by using inference based on the corporate structure, and (b) we do not rate all documents as equal for evidence.

In general, we calculate a person's expertise by analysing the documents that contain that person's name and documents that may not contain their name but may occur "close" to other important documents in some organizational structure. For example, on a corporate intranet we may designate a project page as highly relevant evidence of the expertise of the project members. Documents close to this project page in the intranet structure may also be counted as relevant even if they do not explicitly mention project members' names. Moreover a home page or a project page will be weighted higher as evidence than, for example, a news page or a document page.

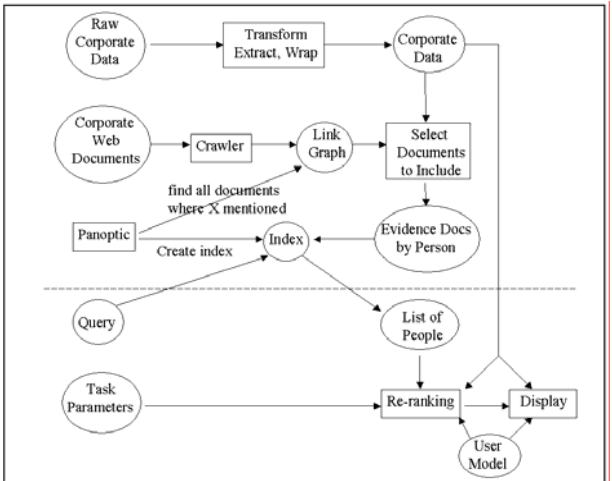

**Figure 1: Data flow architecture diagram**

The architecture of our PeopleFinder system is shown in Figure 1. The top half of the figure represents our offline evidence capture and indexing while the bottom half represents run-time operation.

Corporate organizational data is used extensively to 1) help determine what documents to use as evidence for expertise, 2) to help determine how to rank the returned list and 3) to help determine how to display the results. For this data, we use a manually created XML document that represents the hierarchy of organizational units, existing projects, current staff and the relationships between them[1].

We also use the organization's web graph to provide some information structure and to identify related information to seed points. Using both sets of information we construct evidence fragments for each person (described in §4) which are then indexed.

At run-time, a user may be involved in a task and we represent this as a set of parameters that feed into our PeopleFinder system so that we can perform different rankings and presentations of the results according to the task. We have implemented the "find an expert to ask for information" task where a required role-type can be requested and in the future would like to extend this to "find a team" task.

As well as using the structured corporate information to assist in the ranking and organization of the results, we can employ a user's profile. For example, a scientist may want to see other scientists first whereas a business client may want to see group leaders first.

Figure 2 shows the result of a query to the system. An expert is show with the evidence list expanded. A second window shows the relationship of this person to others in the organization.

---

[1] In a real system, we would expect this information to be constructed automatically from existing corporate data.

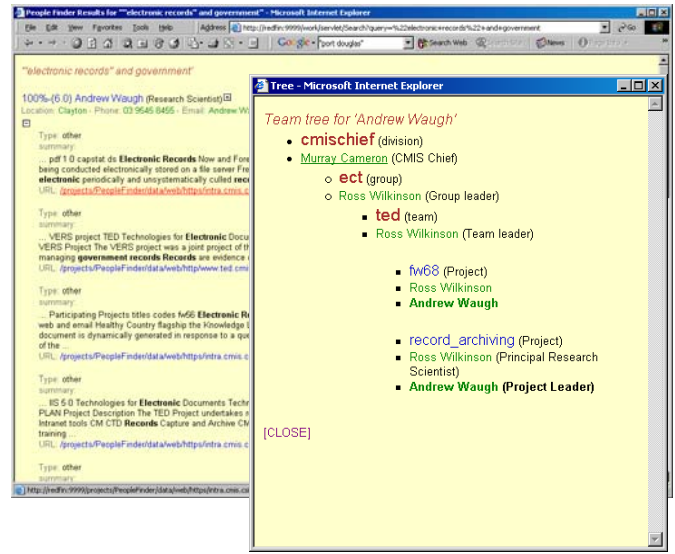

**Figure 2: This screen dump shows the results of a query.**

## 4. BUILDING THE EVIDENCE

Our algorithm for selecting the documents to use as evidence on expertise is as follows:

1. Crawl the web data sources using our Panoptic Funnelback crawler. For our experiments, we used our divisional intranet and extranet. Other corporate data such as a database of contact reports, a database of publication citations, and a database of project reports (referred to below as *corporate database data*) is optionally added to the collection. The collection is then indexed.

2. Construct a web graph from the crawl. We process the log file produced by the crawler to generate a graph structure, taking into account re-directs and aliases. In our experiments this graph has 59760 nodes and 515024 edges.

3. Parse the organizational information structure and extract seed points. The following shows an extract of our organizational information:

```
<unit id="ted" type="team">
    <details>
        <title>…</title>
        <description></description>
        <descriptionurls>
            <url>http://… </url>
        </descriptionurls>
        <member personID="p1">
            <role roleID="tl"/>
        </member>
        <project projectID="expert" />
        ...
    </details>
</unit>
```

Following this we have definitions of projects and people. A project also specifies who is working on that project. Thus for each person, we can extract a homepage and identify a number of relationships such as project and group membership. Both groups and projects may have homepages. We also combine project descriptions that we obtain from the project report database.

4. Construct distance from seed points to other documents in the graph.

We analyse the web graph and use a shortest path algorithm to construct for each persons set of seed points (e.g. a user's home page, a project home page) a distance to all other pages below a threshold. This first step simply counts the number of links from one page to another (see Figure 3). The final weight of a document is proportional to the reciprocal of this number. All these documents are assigned as potential evidence for that person's expertise.

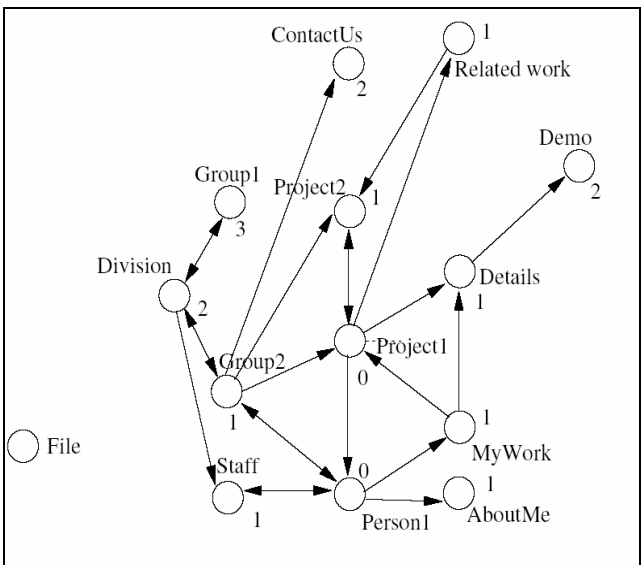

**Figure 3: Initial weights are the lowest number of links from a seed page. In this example seed pages are "person1" and "project1".**

5. Extract document fragments within a certain radius and weight according to location and distance.

In general, the level of importance of a page for a particular user then depends upon the distance, in terms of links, of the page from a "seed" page, *and* the number of levels separating the page from the level of a "seed" page.

We take advantage of the directory structure and so links that connect a page to a page at the same level or below (i.e. something more specific) are stronger (reduce evidence contribution by factor of 2) than links to pages above or away from the current node (reduce by factor of 10). In Figure 4 we have saved some space by grouping a folder's default file (e.g. index.html) with the folder name. So, for example, folder "Project1" contains Project1.html, Details.html, Related Work.html and the folder Demo.

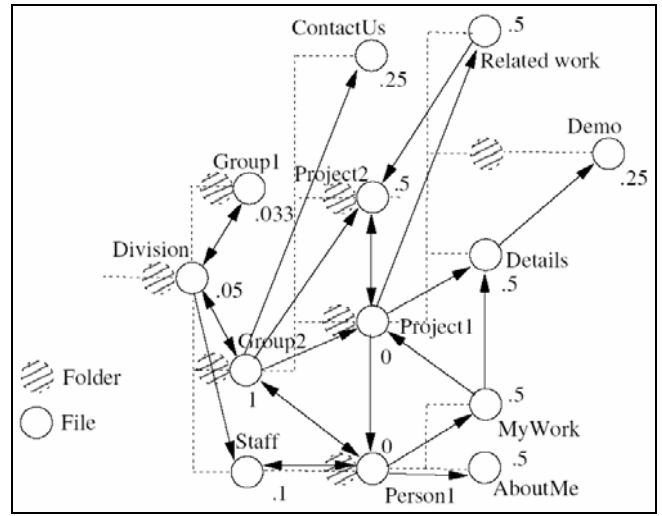

**Figure 4: Assigning weights according to directory hierarchy**

6. Extract all document fragments containing a person's name.

Now we extract all document fragments from the set of crawled pages that contain the person's name. We currently use the entire document content although a window around that person's name occurrence(s) in the document could also be used.

7. Create set of evidence for each person.

The importance of each page in the evidence set is multiplied by a factor associated to its type. Some pages attract particular importance such as home pages, project home pages, and unit home pages. In our algorithm homepages, project homepages and group homepages have type factors of 10. Any other page has a type factor of 1. These values are somewhat arbitrary and we intend to evaluate the impact on changing the value of those factors in future work. The final fragment weight is the fragment weight multiplied by the type factor (Figure 5).

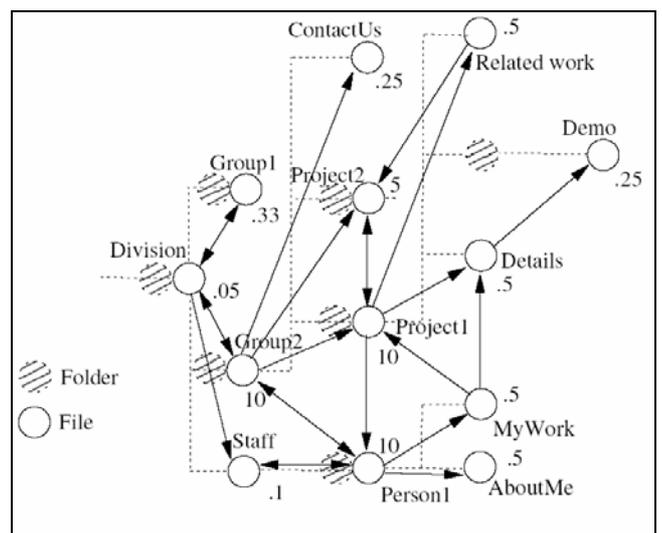

**Figure 5: Final weights assigned to the nodes**

8. Index collection.

The entire collection is then indexed. When a query is issued matching document fragments are returned and are assigned a score which is obtained by multiplying the modified Okapi similarity measure between the fragment and the query multiplied with the final weight of the fragment. For each person the result score is computed by summing up the scores of the individual's evidence fragments.

# 5. EVALUATION

The objective of the evaluations is to be able to answer the following questions:

1. How effective is the new prototype in finding experts? In particular, we wanted to investigate whether the effort of collecting structured organizational data and making use of it to build evidence around each person pays off.

2. What amount and type of corporate data make a difference? In some organizations corporate data may not be easily available, or costly to configure in a format usable by our system. We wanted to investigate in what circumstances the new prototype could achieve its best performance, so that we can save (both human's and machine's) effort on collecting unnecessary documents, and we are confident that the performance of the prototype would not deteriorate if we don't have certain types of documents.

## 5.1 Test Collection

In order to evaluate our PeopleFinder prototype we first need to build a test collection. This test collection consists of a set of documents to search on, a set of topics/queries to search for, and relevance judgments – in our case these are which people in the organization have expertise on each search topic.

Our test collection contains:

1. Web pages from the CMIS division's Extranet and Intranet, collected by our in-house Panoptic crawler, on 20th March 2003. (this collection will be referred to as "web collection".)

2. Organizational information such as staff list and project membership and organizational hierarchy created manually on 20th March 2003.

3. Current project list and members from our corporate project database, for the month of February 2003, as well as current project description from the project plans.

4. Publication list from the CMIS publication page and groups' publication pages.

5. Business development contact database (emails)

Data 3-5 have been manually or semi-automatically created and converted into XML documents and will be referred to as "corporate database data".

A list of 138 test queries has been manually assembled by looking at terms in research group web pages, terms from the ACM thesaurus, and selecting terms from the query log of our previous prototype. Examples of queries are: natural language technologies, mathematical morphology, sampling of minerals, XML protocols, audio analysis, SVG, RDF, data mining, atmospheric science. Here, we are concentrating on topic queries but it is worth noting that in a practical system expertise selection should be augmented by criteria such as preference for a local expert rather than a distant one, estimated load of the expert, willingness or capability in providing assistance. Such criteria could be used to filter or group the results based on expertise.

For the relevance judgments, we first took the top ten people per topic of each run and sent each person a list of topics in which their names were retrieved and asked these people to rank their level of expertise for each topic on four scales: high, medium, low, or none. For those people who did not reply, we used third party (colleagues or group leaders) judgments.

### 5.1.1 Evaluation Criteria

In PeopleFinder we are interested in looking only at a few results, since the purpose is to contact one person. Thus our evaluation is based on precision where we define precision as the percentage of the correctly identified (un-repeated) experts in the retrieved set. Also, the list of assessed experts for most topics is small since the research groups are relatively small[2]. Therefore, we use the average precision over all test queries at cut-off of top ranks: 1, 3, 5 and 10.

The next step is to define the notion of relevance for a given expert. We have defined the following relevance function to map expertise into a relevance value:

if expertise is high or medium, or expertise is low and there are no high/medium level experts available then

$$f_{topic}(person) = 1$$

else

$$f_{topic}(person) = 0$$

Other functions are possible such taking into account only high expertise. Indeed expertise is not absolute; it may depend on how you evaluate yourself, or on the environment you are in. For example, a person with a low level of expertise knows at least what the topic is about and can refer you to better experts. However, since we are only interested in comparing systems rather than the absolute value of the precision, all experiments have the same bias.

Given the complete set of assessments and a quantization function for mapping the assessments to a single relevance value we are able to apply evaluation metrics as in standard document retrieval.

---

[2] For technical expertise, such as java, XML, RDF, C++, OO programming, EJB, etc., the number of experts is much bigger as they can be found in many different groups. We may consider precision/recall @20 in further experimentations for those topics.

### 5.1.2 Experimental Results

We considered two variants according to our testing goals: algorithms and data.

**Algorithms:** First we compared the new prototype (referred to as new system) with our initial system (referred to as base system) to see the benefit of using corporate structure and weighted evidence.

**Data:** Second, we compared using different data sources for evidence of expertise. We are interested in selecting a collection(s) with minimum effort to build while achieving best precision. In this experiment, we used the four document collections: intranet, extranet, web (intranet + extranet), and web plus corporate database data (i.e.: web+db).

We had seven runs in total. The precision for each run is shown in Table 1. We can see that the algorithm outperforms our basic system for any set of data at any cut-off.

**Table 1. The average precision for all runs.**

| p@ | 1 | 3 | 5 | 10 |
|---|---|---|---|---|
| Base-intranet | 0.399 | 0.365 | 0.330 | 0.267 |
| Base-extranet | 0.558 | 0.498 | 0.430 | 0.334 |
| Base-web | 0.435 | 0.391 | 0.355 | 0.301 |
| New-extranet | 0.609 | 0.548 | 0.509 | 0.413 |
| New-extranet+db | 0.659 | 0.556 | 0.517 | 0.414 |
| New-web | 0.616 | 0.556 | 0.484 | 0.372 |
| New-web+db | 0.659 | 0.592 | 0.523 | 0.420 |

In the next two sections we discuss each variant in more details.

### 5.1.2.1 Impact of the new algorithm

Tables 1 and 2 show the comparison of the new system and the base system with the extranet collection and the web collection respectively. Each entry has two numbers, fraction and percentage. In the fraction X/Y: X is the percentage of queries where the run of that row is better than the run of that column. Y is the percentage of queries where the run of the column is better than the run of the row. The percentage number is the overall improvement of the run of the row over the run of the column. The number in bold represents the significant improvement using the paired, two tailed t-test. We can see that the performance of the new system is significantly improved over the base system by using either the extranet collection (18%) or the web collection (36%), this performance is further improved by adding corporate database data into the either the extranet collection (22%) or the web collection (47%).

**Table 2. System comparison with the extranet collection**

| p@5 | Base-extranet |
|---|---|
| New-extranet | **40/20  18%** |
| New-extranet+db | **47/22  22%** |

**Table 3. System comparison with the web collection**

| p@5 | Base-web |
|---|---|
| New-web | **51/22  36%** |
| New-web+db | **56/12  47%** |

### 5.1.2.2 Impact of the data Collections

**Base System**

The base system was tested against three collections: intranet (base-intranet), extranet (base-extranet), and web (base-web). As shown in Table 1, overall, the base system works best with the extranet collection and worst with the intranet collection. Table 4 shows that, at cut-off 5, the base-extranet is significantly better than the base-intranet and the base-web, while there is no significant difference between the base-intranet and the base-web. It is not surprising that the intranet collection is the worst as, in our opinion, our intranet collection does not contain rich sources of expertise information, i.e. home pages and group pages are on the organization's extranet.

**Table 4. Comparison of the base system on three collections.**

| p@5 | Base-intranet | Base-extranet |
|---|---|---|
| Base-extranet | **46/23  30%** | |
| Base-web | 22/15  6% | **21/38  -19%** |

**New System**

The new system was tested against four collections: extranet, extranet+db, web, and web+db. Table 1 shows that, among these four runs, the use of web+db performs the best, closely followed by extranet+db.

We examined the scores for each rank over all queries for both the web and web+db data and found that adding the corporate database data only had an effect on individuals who were ranked between 5 and 12. Thus we see that the corporate database data is capturing important information for some candidates that was not captured by the web data. Looking at individual results we see that if the person had a poor web presence then the corporate database data could increase his score to bring him into the top 10, but with the this was not enough to replace the candidates that scored very highly with the web data (i.e. had relevant home or project pages). Thus we see a slight improvement @1, @3 but significant improvement @5 and @10. The extranet data in our collection is relatively rich and the corporate database data is relatively poor. However, this may be different for other organizations, so the relative importance placed on the different data sources will depend on each organizational environment.

Table 5 shows that significant differences exist only between extranet+db and web, and between web and web+db. It is interesting to see that adding corporate database data to the extranet collection does not make difference, while adding XML to the web collection does make a difference. Again, we see that our intranet is simply not adding any value. In fact the precision @10 shows a significant difference in quality between web and extranet. Our intranet data contains many documents about organizational policies, minutes of managerial meetings etc. These pages contain many names and little technical content and effectively add noise to the evidence. We can postulate a similar line of reasoning to the explanation given for the web+db vs web case.

**Table 5. Comparison of the new system on four collections**

| p@5 | Extranet | Extranet+db | Web |
|---|---|---|---|
| Extranet+db | 16/11 2% | | |
| Web | 22/30 -5% | **22/32 –6%** | |
| Web+db | 33/26 3% | 25/23 1% | **36/19 8%** |

## 6. Summary and Future Work

In this paper we have presented an approach that combines structured corporate data and content for the People Finding task, and an evaluation of the approach. Our initial results are promising and indeed show that the use of structured corporate information improves the precision of finding experts.

We presented our system-oriented evaluation in this paper. We are planning to conduct further evaluations of the system including the impact of different weights for different types of documents and a user evaluation to determine how users interact with the system and its usability.

We can conclude that Expert Finding in not a "one engine fits all" solution. It really requires an analysis of the information space into which a solution will be placed and the appropriate selection and weighting scheme of the data sources.

Finally, we hope to extend the prototype for retrieving teams of people, where roles and positions (scientist, software engineer, business developer) can also be taken in account.

## 7. ACKNOWLEDGMENTS

We wish to thank Shaminda Samaratunge for implementing the user interface and the evaluation programs.